\documentclass[12pt]{elsarticle}

\usepackage{graphicx}
\usepackage{amssymb}
\usepackage{subcaption}
\usepackage{xcolor}
\PassOptionsToPackage{hyphens}{url}
\usepackage{hyperref}
\usepackage[absolute,overlay]{textpos}

\journal{}

\makeatletter
\def\ps@pprintTitle{%
 \let\@oddhead\@empty
 \let\@evenhead\@empty
 \def\@oddfoot{}%
 \let\@evenfoot\@oddfoot}
\makeatother

\begin{document}

\begin{frontmatter}

\title{Campus Traffic and e-Learning during COVID-19 Pandemic}

\author{
    Thomas Favale$^\dagger$, 
    Francesca Soro$^\dagger$, 
    Martino Trevisan$^\dagger$, \\
    Idilio Drago $^\ddagger$,
    Marco Mellia$^\dagger$,
    \\
    \small{$^\dagger$Politecnico di Torino,} \texttt{first.last@polito.it}
    \\
    \small{$^\ddagger$University of Turin,} \texttt{idilio.drago@unito.it} 
}

\begin{abstract}
The COVID-19 pandemic led to the adoption of severe measures to counteract the spread of the infection. Social distancing and lockdown measures modified people's habits, while the Internet gained a major role in supporting remote working, e-teaching, online collaboration, gaming, video streaming, etc. All these sudden changes put unprecedented stress on the network.

In this paper, we analyze the impact of the lockdown enforcement on the Politecnico di Torino campus network. Right after the school shutdown on the $25^{th}$ of February, PoliTO deployed its own in-house solution for virtual teaching.  Ever since, the university provides about $600$ virtual classes daily, serving more than 16\,000 students per day.
Here, we report a picture of how the pandemic changed PoliTO's network traffic. We first focus on the usage of remote working and collaboration platforms. Given the peculiarity of PoliTO online teaching solution that is hosted in-house, we drill down on the traffic, characterizing both the audience and the network footprint. 
Overall, we present a snapshot of the abrupt changes seen on campus traffic due to COVID-19, and testify how the Internet has proved robust to successfully cope with challenges while maintaining the university operations.


\end{abstract}


\end{frontmatter}

\TPshowboxestrue
\TPMargin{0.3cm}
\begin{textblock*}{15.5cm}(3cm,1.4cm)
\footnotesize
\bf
\definecolor{myRed}{rgb}{0.55,0,0}
\color{myRed}
\noindent
Please cite this article as: Thomas Favale, Francesca Soro, Martino Trevisan, Idilio Drago, Marco Mellia, Campus Traffic and e-Learning during COVID-19 Pandemic, Computer Networks (2020), DOI: \url{https://doi.org/10.1016/j.comnet.2020.107290}
\end{textblock*}

\section{Introduction}
\label{sec:intro}

Since its first outbreak between late 2019 and early 2020 in China, the COVID-19 pandemic has had a massive impact on people's lives and habits. The countries most affected by the virus spreading are facing an unprecedented health crisis, whose effects will impact their economic and social structures for a long time. The urge to respect social distancing and lockdown measures adopted to limit the spreading of the infection led to a shift in the fruition and supply of a wide number of services. Some examples include the increased usage of home delivery services, the shift to online lessons and the adoption of remote working solutions. 

Italy has been among the first countries hit by COVID-19. The first case was identified in the north of Italy on February 21$^{st}$, and on the same date, the Government issued the fist law decree to impose quarantine in small selected towns.
On February 25$^{th}$, the Government extended the restrictions to impose remote working for all public offices, shutting down schools, and classes at Universities. Restrictions applied to the largest four regions in the north of Italy.
On March 1$^{st}$, these restrictions were extended to the whole Italy, with further lockdown actions entering in place on March 4$^{th}$ and 8$^{th}$.
On March 11$^{th}$, the ``\#IoRestoACasa'' Prime Ministerial Decree imposed a total lockdown to the whole Italy. Since then, and still at the time of writing, people are allowed to exit from home only for specific and urgent needs. Common retail businesses, catering and restaurant services are suspended. Gatherings in public places are prohibited. All activities not deemed essential for the Italian production chain are closed. Italy entered the most restrictive lockdown in its history.

\begin{figure}[t]
    \begin{center}
        \includegraphics[width=0.6\columnwidth]{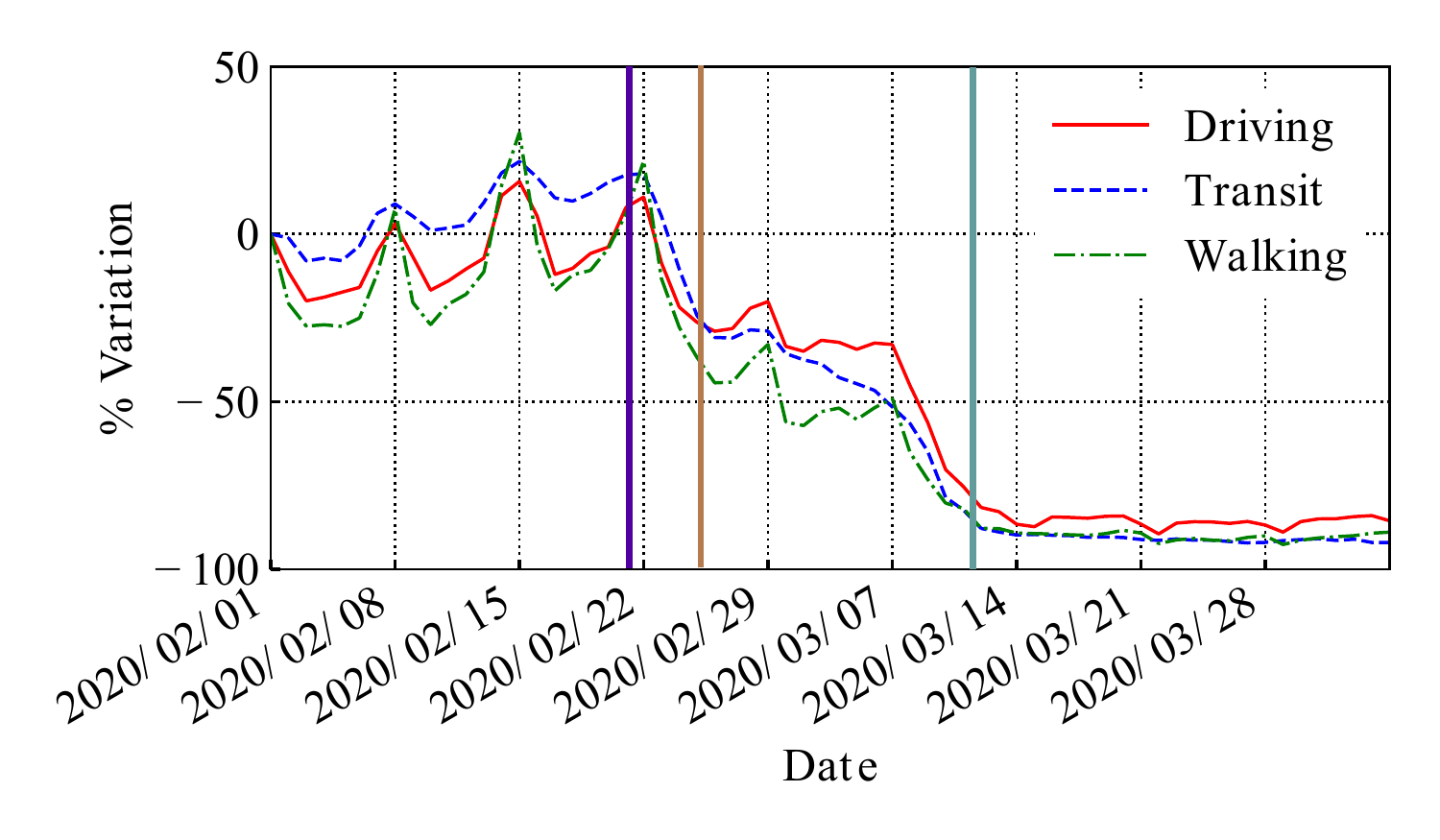}
        \caption{Variations in the mobility patterns in Italy between February and April 2020. Colors mark the days in which new Ministerial Decrees introduced mobility restrictions. Source: \url{https://www.apple.com/covid19/mobility}.}
        \label{fig:mobility}
    \end{center}
\end{figure}

Figure~\ref{fig:mobility} clearly depicts the impact of these measures.\footnote{Source: \url{https://www.apple.com/covid19/mobility}} It shows the variation of the mobility patterns in Italy since the begin of the emergency and the impact of the restrictions and total lockdown. The vertical bars highlight the relevant events listed above. The leftmost violet bar identifies the date of the first COVID-19 case in Italy. The central brown bar identifies the school shutdown. The rightmost bar marks the date of the ``\#IoRestoACasa'' decree. The decrease in mobility is drastic after each of those events.

Restrictions limited people's mobility while remote working, e-learning, online collaboration platforms started to grow along with online leisure solutions, like gaming and video streaming. These new habits highlighted the fundamental role of the Internet. Correspondingly, Internet traffic volume has grown by about $40\%$, sometimes with a decrease in the download performance, questioning the resiliency of the Internet itself~\cite{cloudflare,fastly}.

In this paper, we analyze the changes in the traffic patterns that are visible from our university campus, the Politecnico di Torino (PoliTO for short) in Italy. We look at the campus traffic, focusing on collaboration and remote working platforms usage, remote teaching adoption, and look for changes in unsolicited/malicious traffic. PoliTO opted to implement an in-house e-learning solution based on the BigBlueButton framework to support all the classes of the second semester, which were scheduled to start on March $2^{nd}$.\footnote{
\url{https://bigbluebutton.org/}} The platform has been designed, installed, and tested during the first week of March, going live for the starting of the online semester one week later.
Here we leverage this unique point of view to observe changes in the campus traffic and services during such a singular event. Moreover, we dig into details on how students access online classes and teaching material. Since students enjoy classes from their homes at different places, connected by different network operators, we check whether and how these factors affect the performance of the online teaching systems.

Overall, we highlight a 10 times decrease in incoming traffic during the lockdown. Outgoing traffic grows instead of 2.5 times, driven by more than 600 daily online classes, with around 16\,000 students per day that follow classes. Online collaboration exploded, with faculty and staff members exchanging more than 17\,000 chat messages and participating to more than 1\,000 calls per day. We observe a surge in remote learning and working also during the weekends. Considering Internet connectivity, we notice no major problems, with only a few cases of poor performance, possibly related to people connected via 3G/4G operators. 

In a nutshell, we believe this paper testifies how the Internet proved able to cope with the sudden need for connectivity. We attest how remote working, e-learning and online collaboration platforms are a viable solution to cope with the social distancing policies during COVID-19 pandemic.
While we all hope the latter will soon be relieved, we believe the experience of online collaboration will continue also after the COVID-19 disappears.

The paper is organized as follows: Section~\ref{sec:dataset} describes the datasets and methodology; Section~\ref{sec:campus_traffic} reports a drill-down on the aggregate campus traffic patterns and online collaboration solutions;  Section~\ref{sec:virtual_teach} focuses on online classes; Section~\ref{sec:net_perf} details network performance metrics, breaking down by operator and region; Section~\ref{sec:unsolicited} looks into possible changes in unsolicited and malicious traffic, such as portscans and spam emails; Section~\ref{sec:relatedwork} summarizes related work; finally Section~\ref{sec:conclusions} concludes the paper.



\section{Datasets and methodology}
\label{sec:dataset}

PoliTO is a medium-sized university that offers bachelor, master and graduate-level courses in the engineering and architecture fields only. It is among the top Universities in Italy. PoliTO has about 35\,000 students, of which 30\% come from the Piedmont region where Torino is, 55\% from other Italian regions, and 15\% from the rest of the world. PoliTO employs about 2\,000 faculty and researchers, and around 1\,000 administrative staff members.\footnote{\url{https://www.polito.it/ateneo/colpodocchio/index.php?lang=en}} PoliTO main campus network hosts all its IT services, and offers both Ethernet and WiFi connectivity to departments, offices, classrooms, student rooms, and laboratories. Two 10\,Gbit/s access links connect the campus LAN to the Internet, through the GARR (Gruppo per l'Armonizzazione delle Reti della Ricerca) network, which provides Internet access to all Italian universities. 

In the following, we describe the data sources we use to gauge the changes in the traffic and services hosted at the campus network. If not explicitly said, the data cover the period from Saturday, February 1$^{st}$ to Sunday, April 5$^{th}$ 2020.

\subsection{Passive traces}

We leverage statistics collected by edge routers to observe and compare the load on different Italian University networks. This data is stored by GARR in a central database publicly accessible.\footnote{\url{https://gins.garr.it/home\_statistics.php}} The repository stores the time series of the traffic volume on each edge link of the GARR network, with a granularity of five minutes. In our analysis, we consider PoliTO campus and compare it with two other large Italian universities for reference, namely Politecnico di Milano (the biggest technical university in Italy) and Università di Torino.\footnote{Università di Torino is a separate University which offers degrees on sciences, humanities, economics, etc.}

While GARR offers only high-level aggregated data, here we also leverage on fine-grained measurements exposed by the monitoring infrastructure deployed in the Campus network. Such infrastructure is based on Tstat~\cite{trevisan2017traffic}, a passive sniffer that analyzes the traffic entering and leaving the Campus. It computes flow-level logs similar to NetFlow, exposing information about TCP and UDP flows observed in the network. Beside classical flow-level fields, such as IP addresses and port numbers, Tstat exposes metrics such as Round-Trip time (RTT) and packet losses. We store Tstat logs in a secured Hadoop-based cluster. Tstat limits as much as possible any information that can be used to identify users -- e.g., client IP addresses are anonymized using the CryptoPan algorithm~\cite{fan2004crypto}, and no payload is stored. All data collection is approved and supervised by the responsible University officers.

\subsection{Application logs}

In addition to passive measurements, we extract data about servers and network devices offering specific services and applications that staff members may use for smart working. We focus on three classes of services: (i) the tools to access the internal resources of the Campus while working from home, namely Virtual Private Network (VPN) and Remote Desktop Protocol (RDP) services; (ii) the Microsoft Teams collaborative platform, which PoliTO adopted since 2019 to offer chat, file sharing, video calls and collaboration services to employees and students; (iii) email, antispam and security services. Logs available on the management consoles offer rich information about the usage of these services over time and let us study how the adoption of such services varied during the epidemic.

\subsection{Virtual classrooms}
\label{sec:dataset_vc}

To face the lockdown during the COVID-19 outbreak, PoliTO opted to set up an internally-hosted virtual classroom service for all classes. The system is based on the BigBlueButton framework for online learning, with customization to integrate it in the existing teaching portal. A total of 
41 high-end servers running the BigBlueButton components and hosted in the Campus data center were set up in the first week of March. All classes started a week late. 
The BigBlueButton client-side application is based on HTML5. It provides high-quality audio, video and screen sharing application using the browser’s built-in support for web real-time communication (WebRTC) libraries.
In a nutshell, once the virtual classroom has been set up, BigBlueButton servers act as WebRTC Selective Forwarding Unit (SFU) capable of receiving multiple media streams (video, audio, screen sharing, etc.) and deciding which of these media streams should be sent to which participants. Video is encoded using the high-quality open-source codec VP8. PoliTO's setup lets the lecturer choose four video quality levels with a bitrate of 50, 100 (the default), 200, 300 kbit/s. Screen sharing may require the highest share of bandwidth, and, if the presenter’s screen is updating frequently, the BigBlueButton client could transmit up to 1.0 Mbit/s. Audio is encoded using the OPUS encoder at 40~kb/s.\footnote{\url{https://docs.bigbluebutton.org/support/faq.html\#bandwidth-requirements}} Finally, sharing slides takes little bandwidth beyond the initial upload/download of the pdf file.

In our analyses, we collect and process the application logs of the video servers to study the consumption and performance of virtual classrooms. We also make use of the logs of the legacy teaching web platforms to study access to the teaching material like lecture notes, pre-recorded classrooms, etc.

\subsection{Security monitors}

At last, we analyze logs from the campus border firewall that limits the access towards authorized servers while blocking possibly malicious traffic and well-known attacks. Since we are interested in remote-working solutions, we check the firewall for alarms related to SSH, RDP, SIP attacks. Intuitively, we want to check if the attack patterns changed during the COVID-19 pandemic.

The firewall is configured to let three /24 subnets to be completely open to the Internet, with no hosts connected to it. These sets of addresses act as ``darknets'', i.e., sets of IP addresses regularly advertised which do not host any client or server. Any traffic the darknets receive is unsolicited by definition~\cite{SoroDarknet2020}. By passively analyzing incoming packets, we observe important security events, such as the appearance and spread of botnets, DDoS attacks using spoofed IP address, etc. We use this information to further quantify if there is any change in attack patterns to our campus network during the pandemic.
 
\section{Impact on campus traffic and remote working solutions}
\label{sec:campus_traffic}

In this section, we analyze the impact of the COVID-19 outbreak on campus networks. We first provide quantitative figures on the overall traffic volume changes. Then, we focus on services supporting smart working.

\subsection{Aggregate traffic volume}

\begin{figure}[t]
    \begin{center}
        \begin{subfigure}{0.32\textwidth}
            \includegraphics[width=\columnwidth]{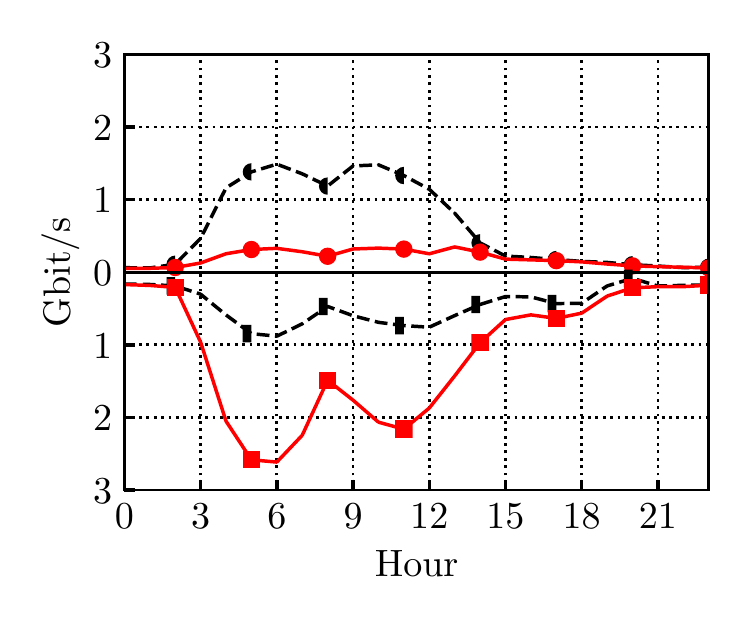}
            \caption{Politecnico di Torino.}
            \label{fig:traffic_volume_polito}
        \end{subfigure}
        \begin{subfigure}{0.32\textwidth}
            \includegraphics[width=\columnwidth]{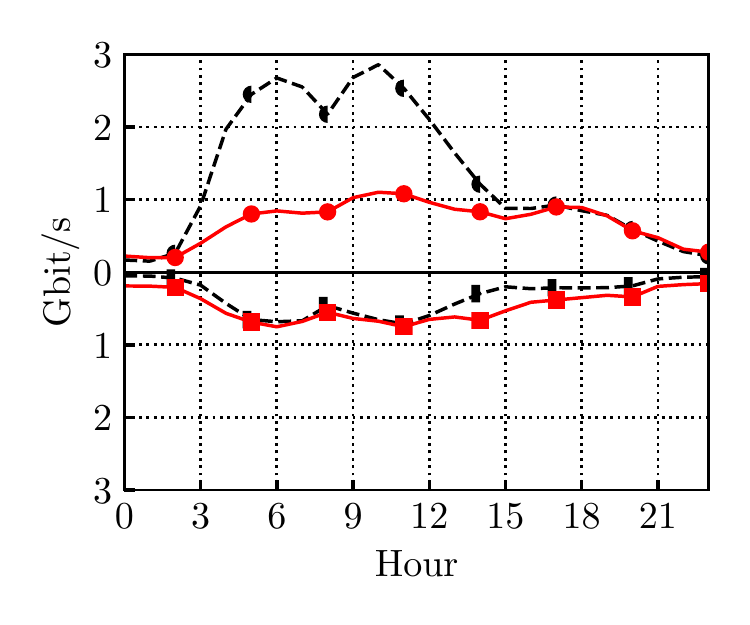}
            \caption{Politecnico di Milano.}
            \label{fig:traffic_volume_polimi}
        \end{subfigure}
        \begin{subfigure}{0.32\textwidth}
            \includegraphics[width=\columnwidth]{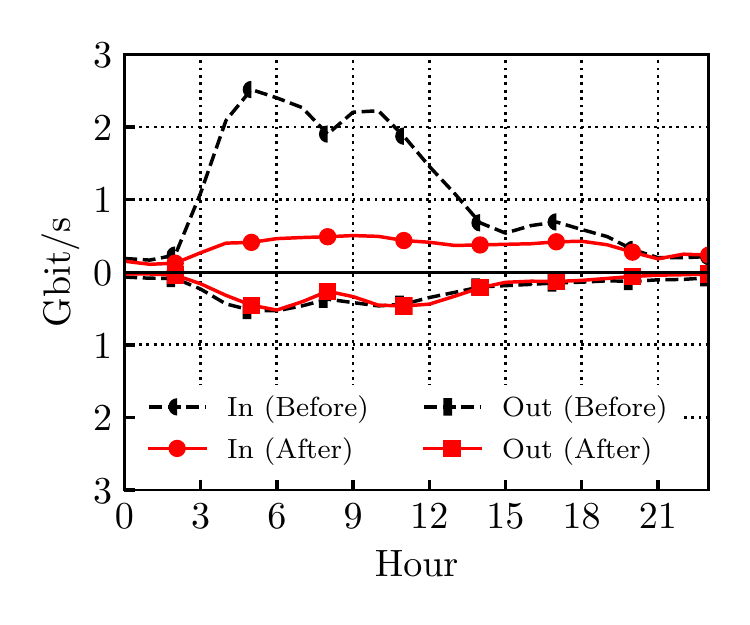}
            \caption{Università di Torino.}
            \label{fig:traffic_volume_unito}
        \end{subfigure}
        \caption{Traffic from three Italian Universities before and after the lockdown.}
        \label{fig:traffic_volume}
    \end{center}
\end{figure}

We first focus on the volume of traffic entering and leaving three Italian university campuses. Figure~\ref{fig:traffic_volume} shows the average hourly bitrate. For each hour, we compute the average bitrate seen over five working days. The positive $y$ values represent incoming traffic (traffic directed to clients and servers hosted in the Campus LAN), while negative $y$ values report the outgoing traffic (traffic directed to clients and servers on the Internet).
Lines mark the volumes before and after the lockdown: the black lines depict the average per-hour bitrate observed during the week before the lockdown (third week of February), the red ones refer to averages calculated on the second week after the lockdown (third week of March).
For comparison, the figure includes plots for the Politecnico di Torino (PoliTO) in Figure~\ref{fig:traffic_volume_polito}, the Politecnico di Milano (PoliMI) in Figure~\ref{fig:traffic_volume_polimi} and the Università di Torino (UniTO) in Figure~\ref{fig:traffic_volume_unito}.

Focusing on the positive $y-$axes, observe how the incoming traffic has shrunk in the three cases, reflecting the lockdown effects. Since the second week of March, most students, researchers and staff members cannot access the campuses. The traffic after the lockdown is about one tenth of the traffic before it in both PoliTO and UniTO, with PoliMI still observing some sizeable incoming traffic. This difference reflects the different lockdown policies imposed by each university. PoliTO and UniTO completely blocked all teaching and research activities, while PoliMI still allows the activity of some laboratories.

The negative $y-$axes report the outgoing traffic. Again, the three campuses present different behaviors: We see a major increase in outgoing traffic from PoliTO, which is not observed in other campuses. This behavior is justified by the online teaching platform hosted in PoliTO which causes an increase of about 2.8 times the baseline outgoing traffic during peak time.
PoliMI and UniTO have resorted to cloud-based solutions, hence the outgoing traffic volume does not show relevant changes before and after the lockdown.

\textbf{Take away:}
\textit{Campus incoming traffic drastically reduced during lockdown. Outgoing traffic changed in PoliTO, where an in-house online teaching service has been deployed. This system caused a massive inversion on traffic patterns, with significant growth in upload traffic due to online teaching services.}

\subsection{Smart working adoption}

We now focus on PoliTO only and drill down on the services used by the personnel for working remotely. 

\begin{figure}[t]
    \begin{center}
        \includegraphics[width=0.7\columnwidth]{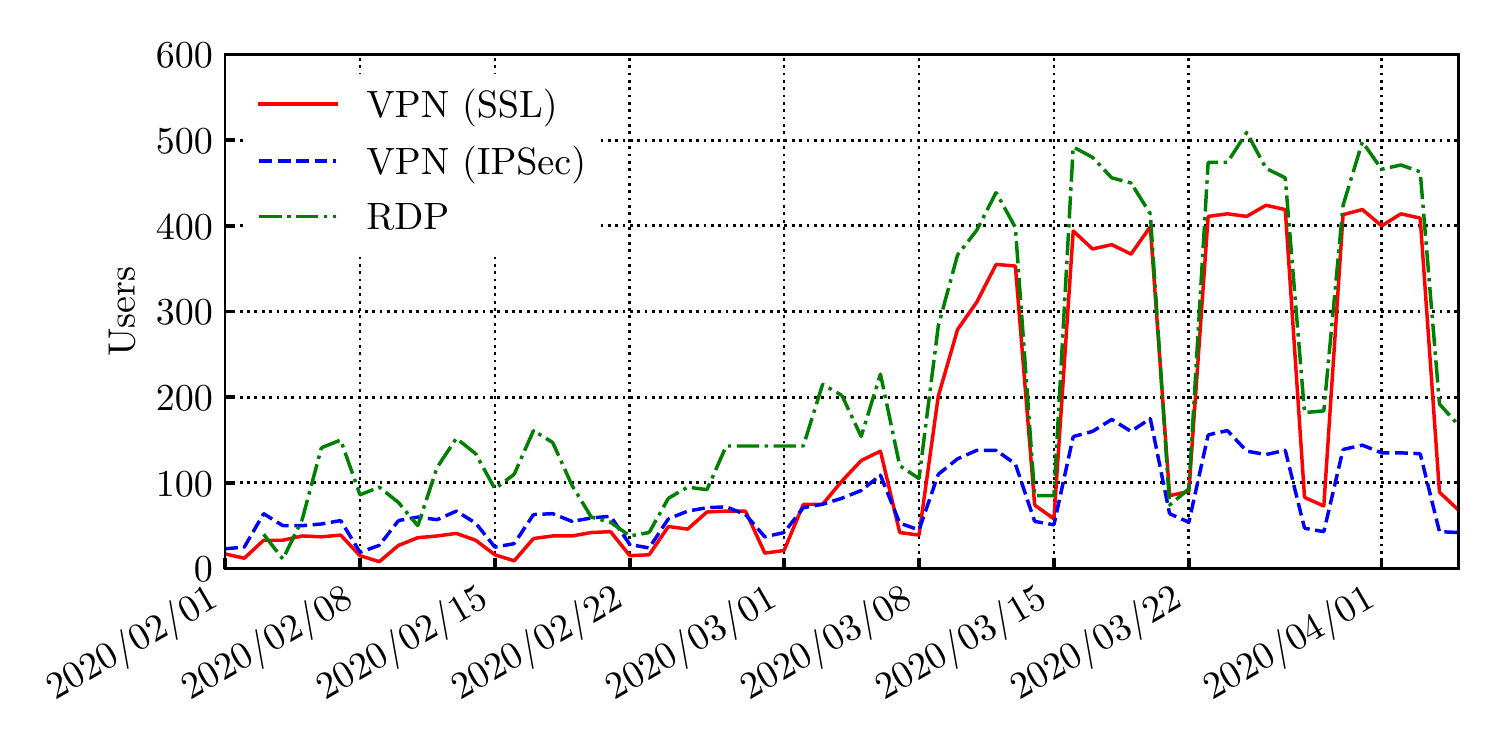}
        \caption{Daily accesses to in-house PoliTO smart working systems.}
        \label{fig:vpn}
    \end{center}
\end{figure}

In Figure~\ref{fig:vpn} we show how the number of users relying on VPNs and remote desktop (RDP) solutions have changed over time. VPNs are used to access in-campus services and servers, while RDP allows one to remotely access files or run applications on their office computers.
Curves in Figure~\ref{fig:vpn} depict the total number of unique users accessing the services at least once in a day.\footnote{VPN accesses are directly extracted from the VPN terminator logs. We identify RDP connections from Tstat logs, considering TCP flows directed to port 3389 that exchanged at least 10~MB of data. We count users by their client IP addresses, which  we assume to be unique on a daily basis.} The effect of the lockdown started on March $11^{th}$ 2020 is astonishing. Since the lockdown started, these solutions simultaneously present a relevant increase in usage. PoliTO offers VPN services both over IPSec and SSL. Interestingly, SSL-based VPN usage increases significantly more than the IPSec-based solution. This suggests that the lockdown forced non-expert users to resort to a VPN, and they have opted for SSL-based VPN, which is easier to configure.

Users working from home also heavily rely on RDP, with about 500 users contacting such services at least once in a day after the lockdown. Sessions (not reported for the sake of brevity) last several hours, suggesting that this remote access method is mostly used for regular working sessions, and not only to access files on computers left in offices. 
Notice also the growth in the number of accesses over weekends, with more than 200 RDP accesses per day. This suggests that people, forced at home, keep working during the weekend, too.

\begin{figure}[t]
    \begin{center}
        \includegraphics[width=0.7\columnwidth]{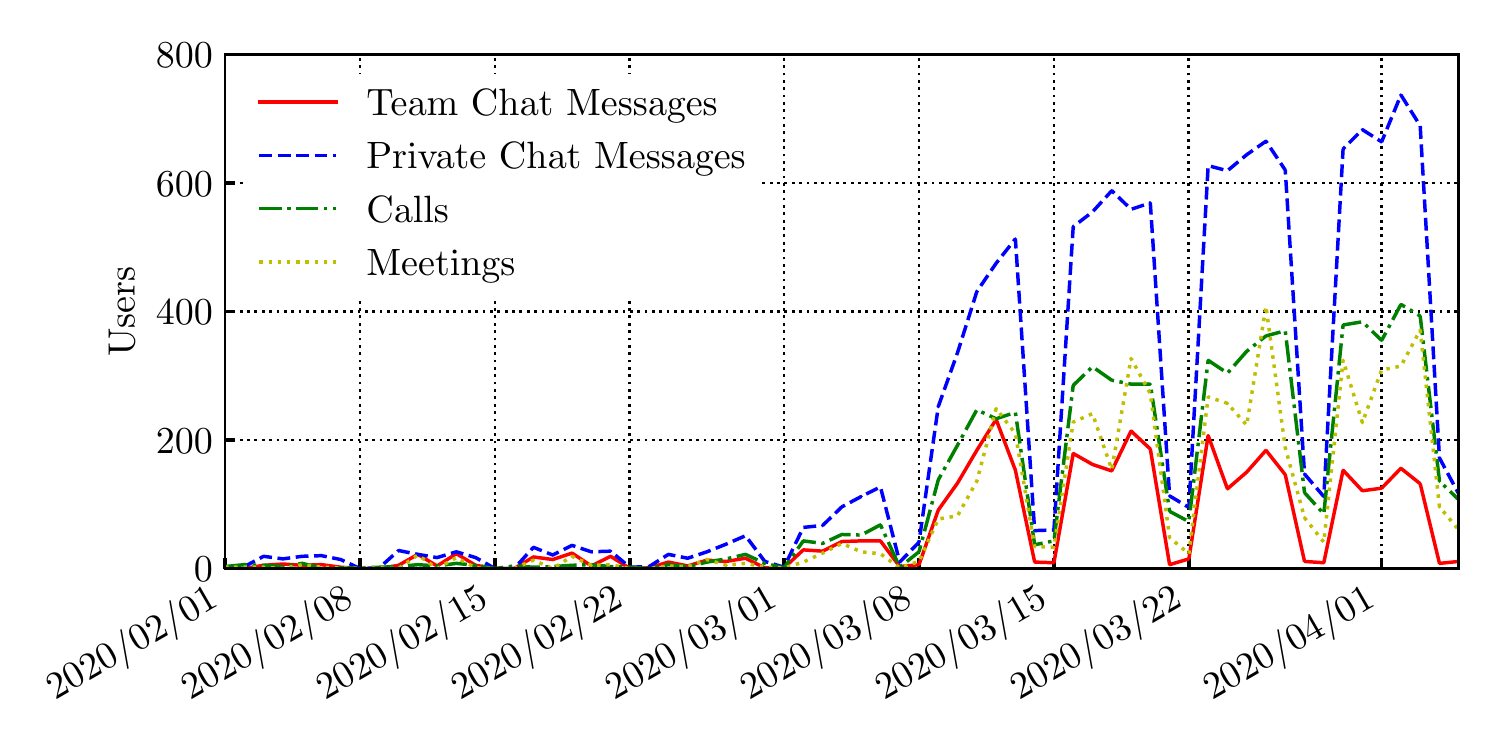}
        \caption{Daily activity on PoliTO Microsoft Teams systems.}
        \label{fig:teams}
    \end{center}
\end{figure}

\subsection{Remote collaboration}

We now move to remote collaboration suites. PoliTO offers all employees and students free access to the Microsoft Teams platform for smart working. Running on the cloud, it puts little loads on the campus network. Yet, its usage provides insights on people's habits during the lockdown.\footnote{PoliTO staff members are not compelled to use Microsoft Teams, we, therefore, expect other platforms to show similar significant growth.} Figure~\ref{fig:teams} summarizes the activity of Microsoft Teams users. It shows the number of \emph{team chat messages} (i.e., chat messages to groups), \emph{private chat messages}, \emph{calls} and \emph{meetings}. In Teams' jargon, both \emph{calls} and \emph{meetings} are online video conferencing; \emph{calls} have two participants.\footnote{\url{https://docs.microsoft.com/it-it/microsoftteams/teams-analytics-and-reports/teams-reporting-reference} - accessed April 2020}

Each point in Figure~\ref{fig:teams} marks the number of people using each functionality at least once in a day. Teams was already available but only marginally used in the campus before the lockdown, with few tens of users per day. As for other services supporting smart working, its usage explodes during March.
Compared to smart working connectivity solutions -- see Figure~\ref{fig:vpn} -- the growth is slightly less abrupt, showing that the people resorted first to means to access their data and office computers, and then to online collaboration tools. People rely on Microsoft Teams as a means to exchange direct messages -- more than 700 daily users sent private chat messages over Teams in the last week of March. In total, they exchanged more than 17\,000 messages per day. Interestingly, video calls and meetings keep growing as well, topping to 400 users per day, making more than 1\,500 calls per day. Team chat messages instead show a drop in popularity after an initial surge. 

\textbf{Take away:}
\textit{The lockdown has pushed smart working to widespread adoption. Remote services access via VPN and RDP, VoIP communications and online conferencing suddenly become regular for PoliTO users. }

\section{Online teaching}
\label{sec:virtual_teach}

In this section, we study in detail the fruition and the performance of the online teaching system deployed at PoliTO. The compelled moving of all classes to an online solution allows us to evaluate (i) the impact of such a scenario on the campus networks, (ii) how students accessed these new services, and (iii) if students suffered any impairment due to limitation in the remote connectivity.

\subsection{Audience}

\begin{figure}[t]
    \begin{center}
        \includegraphics[width=0.7\columnwidth]{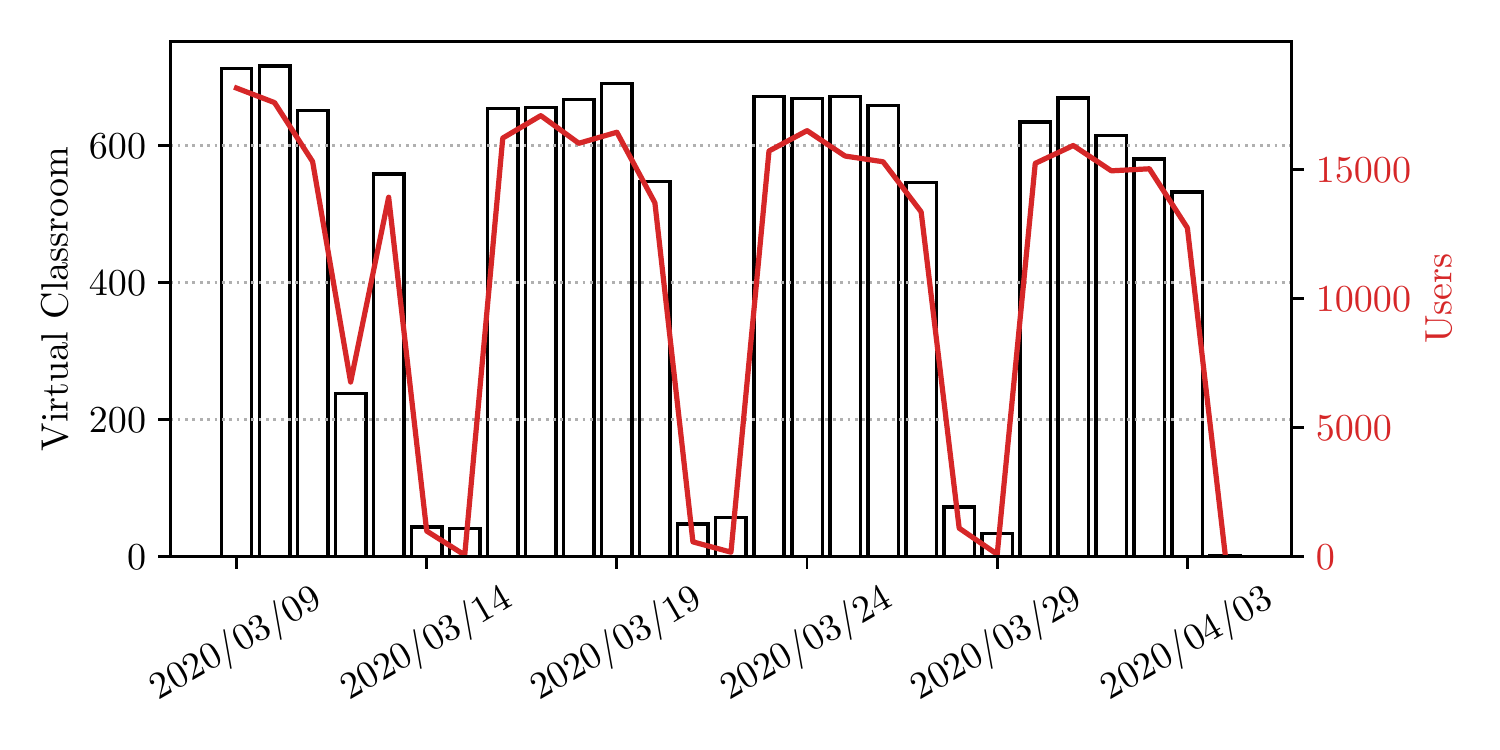}
        \caption{Daily number of virtual classrooms and connected students (faculties and students).}
        \label{fig:virtual_classroom}
    \end{center}
\end{figure}

Figure~\ref{fig:virtual_classroom} provides a summary of the audience of PoliTO's online teaching system. Only March is shown since the university has deployed the system from scratch to cope with the emergency.
 
The number of virtual classrooms (black bars) follows the pattern of the university lecturing schedules, with around 700 virtual classrooms per day. The drop registered on Thursday of the first week was caused by an outage. More than 16\,000 students connect to at least one virtual classroom on a daily basis. Considering that around 35\,000 students are registered at the university, and some join a few lectures a day, a large percentage of the community engages in online teaching each day. The trend on active classrooms and students is stable over time, suggesting that students and faculty found the virtual teaching platform appropriate. 
This is confirmed by the feedback students leave at the end of each class which sees more than 70\% of students giving a feedback of 4 or 5 stars (5 being the highest rating) to the technical quality of the session.

Interestingly, 47\% of classes have been followed by students outside Piedmont, where PoliTO resides. Some students have indeed returned to their home countries or regions before the start of the second semester and could not return to Torino because of the lockdown. Other statistics show that most students follow classes using a Windows 10 PC (64.5\%) or Mac OS X (14.1\%), and Chrome (71\%), Safari (10.1\%) or Firefox (9.9\%) browser.
Yet, 6.2\% and 2.4\% of students follow the class from an Android or iOS device, respectively. 

\textbf{Take away:}
\textit{Engagement in online teaching is impressively high during the COVID-19 emergency, with a large part of the academic community participating on a daily basis. Virtual classrooms allowed students to follow classes, with almost 50\% of them connecting from outside Piedmont.}

\subsection{Network workload}

How much massive online teaching costs in terms of network traffic? Clearly, the answer to this question depends on the technical parameters of the online teaching environment. PoliTO's infrastructure includes the deployment of teaching tools for (i) virtual classroom; (ii) on-demand video of pre-recorded classes (stored as 720p video); (iii) other teaching material (e.g., slides, teachers' notes, other useful files). The virtual classroom service allows the students to watch a live video of the lecturer, an optional whiteboard as well as the direct sharing of the lecturer's screen. The BBB platform uses dedicated servers and WebRTC to set up multiple RTP streams that carry the live video and audio. On-demand video and teaching material are instead offered by additional servers as standard HTTP downloads. 

\begin{figure}[t]
    \begin{center}
        \begin{subfigure}{0.65\textwidth}
            \includegraphics[width=\columnwidth]{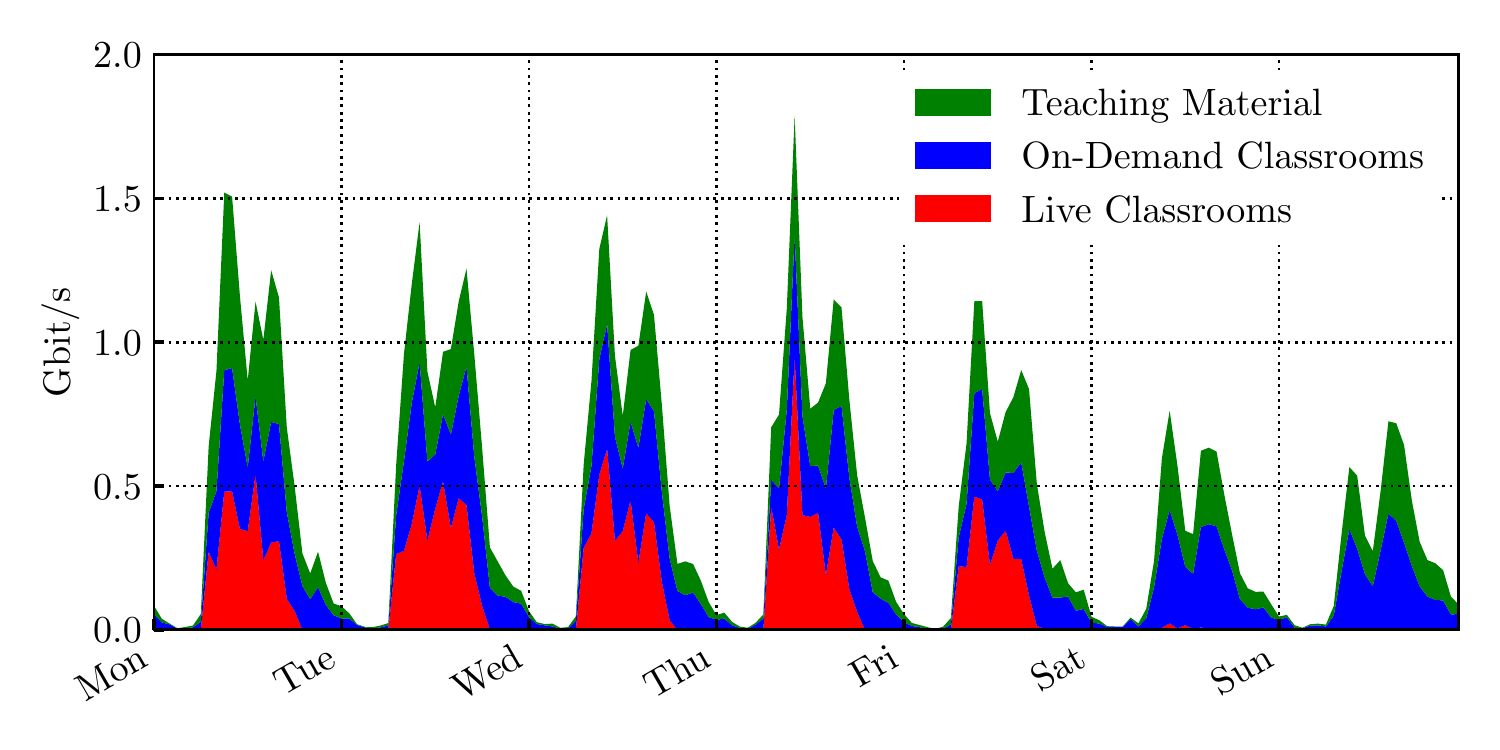}
            \caption{Weekly.}
            \label{fig:learning_pattern_week}
        \end{subfigure}
        \begin{subfigure}{0.65\textwidth}
            \includegraphics[width=\columnwidth]{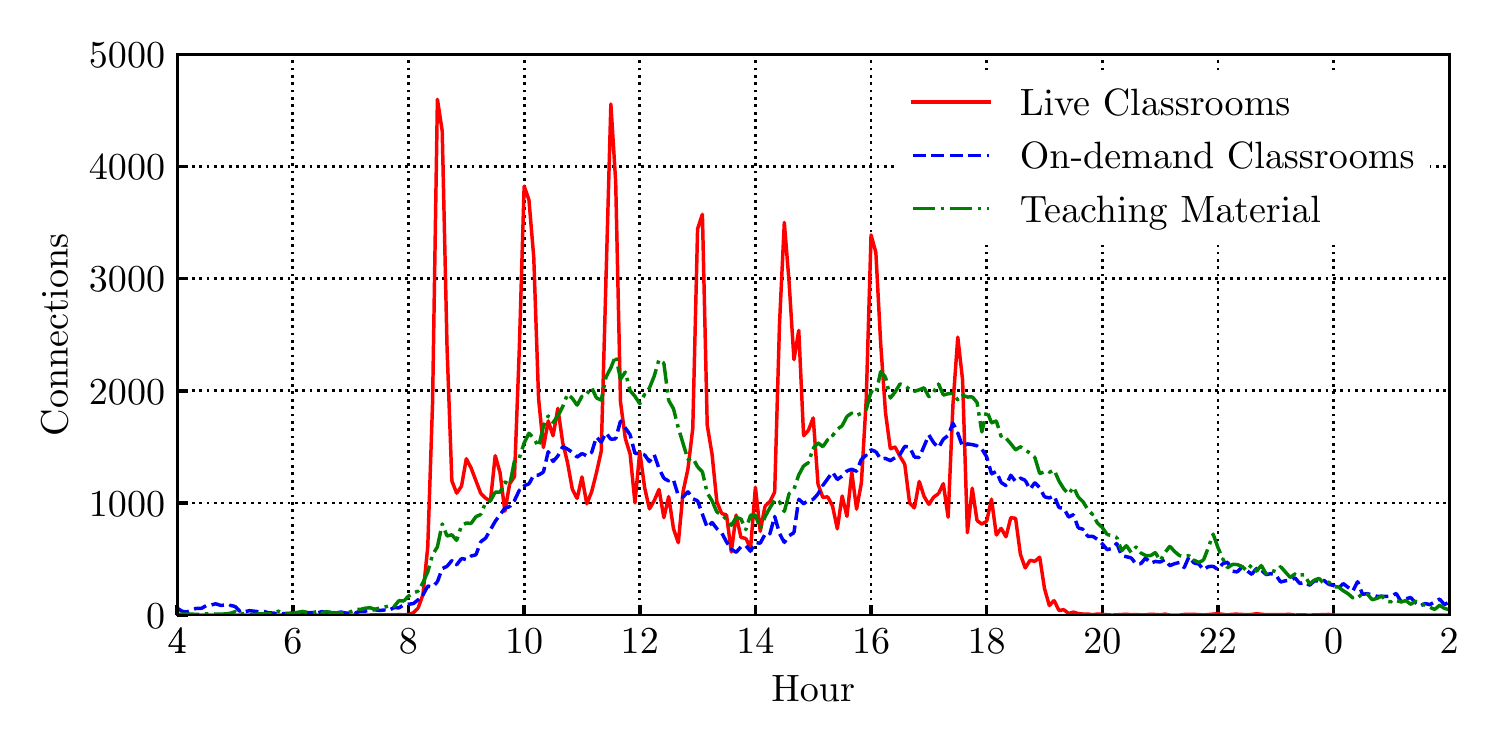}
            \caption{Daily.}
            \label{fig:learning_pattern_hour}
        \end{subfigure}
        \caption{Access pattern to the teaching services.}
        \label{fig:learning_pattern}
    \end{center}
\end{figure}

Figure~\ref{fig:learning_pattern_week} details the traffic volume for each service during the fourth week of March. The figure reports the average bitrate of the traffic leaving the Campus -- i.e., from the servers to the clients, for each hour. During weekdays, the total bitrate exceeds 1~Gbit/s, with live streaming of classes responsible for a bit more than one third of the traffic. During the weekend, when no lecture is scheduled, we still observe large traffic (up to 750~Mbit/s) due to students downloading on-demand lectures and teaching material.
Virtual classroom traffic starts and ends within the schedule, while students keep accessing on-demand classroom and material also at night.

Figure~\ref{fig:learning_pattern_hour} shows how the accesses to teaching facilities are distributed over the day. We take the start time of each student connection to a teaching server and plot the number of new connections observed every five minutes. Accesses to live classrooms (red solid line) clearly follow the schedule of the campus lectures, which begin every 90 minutes, from 8:30 AM until 7 PM. When lectures begin, thousands of students start a new session with the live classrooms servers, resulting in peaks of more than 4\,500 connections. Different is the case for on-demand classrooms and teaching material (blue and green dashed lines, respectively), whose consumption is spread over the day. As said above, students download such material also late in the evenings, including sizeable accesses even after midnight.

\begin{figure}[t]
    \begin{center}
        \begin{subfigure}{0.45\textwidth}
            \includegraphics[width=\columnwidth]{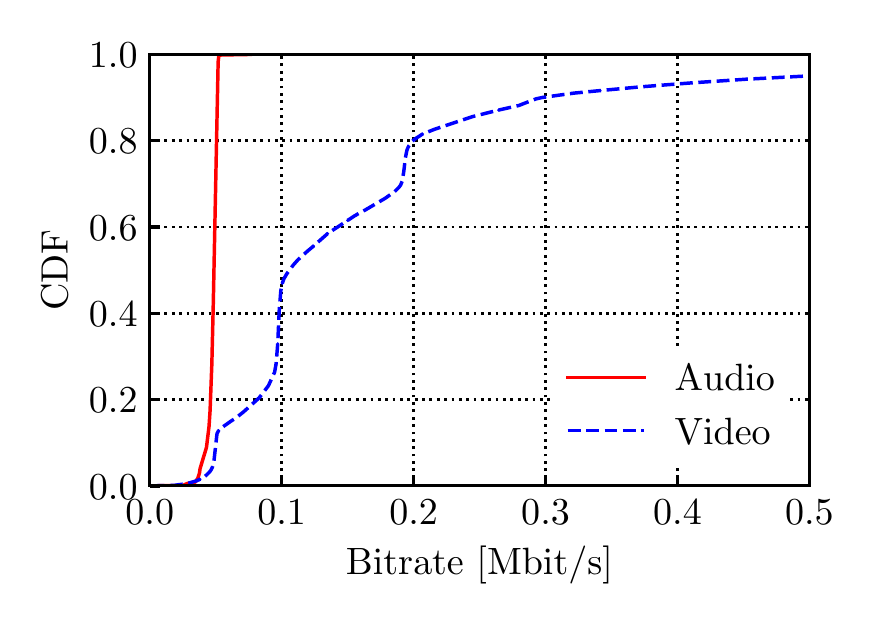}
            \caption{Bitrate.}
            \label{fig:virtual_classroom_bitrate}
        \end{subfigure}
        \begin{subfigure}{0.45\textwidth}
            \includegraphics[width=\columnwidth]{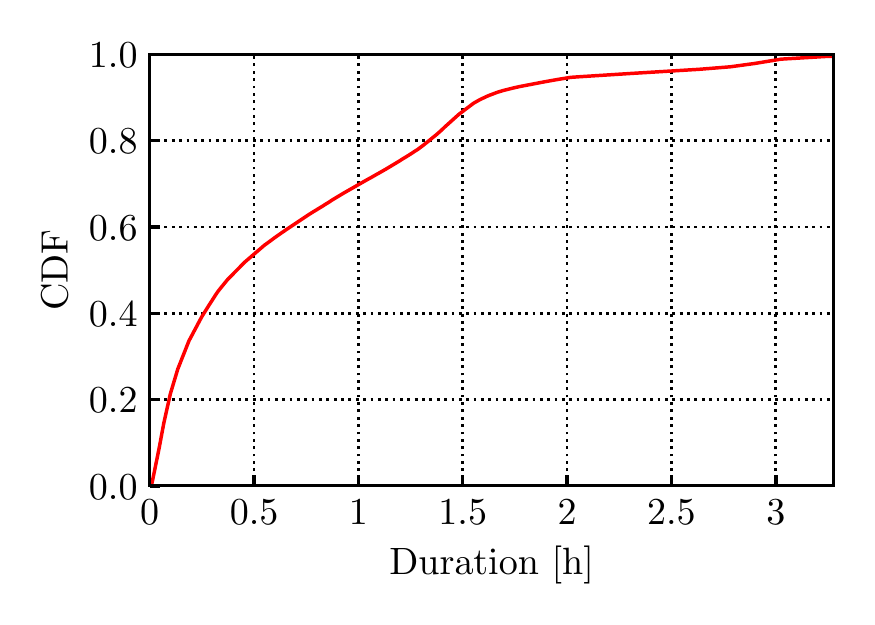}
            \caption{Duration.}
            \label{fig:virtual_classroom_duration}
        \end{subfigure}
        \caption{Characteristics of virtual classroom sessions.}
        \label{fig:virtual_classroom_charact}
    \end{center}
\end{figure}

We complement the above picture with Figure~\ref{fig:virtual_classroom_charact}, in which we show different characteristics of the virtual classroom sessions. Using logs exported by Tstat, we analyze the RTP streams that serve multimedia content (audio and video) to the students and report the distribution of the per-flow average bitrate in Figure~\ref{fig:virtual_classroom_bitrate}, separately per audio and video.\footnote{RTP streams are defined by the combination of endpoint addresses and port numbers, plus the SSRC stream identifier. We distinguish audio and video using the RTP Payload Type field.} All classes carry an audio track, with 40~kbit/s average bitrate -- the expected VoIP bitrate according to the BBB configuration. Considering video, we observe that lecturers usually select the medium quality at 100~kbit/s, but we observe a considerable amount of streams at 50~kbit/s (low quality) and 200~kbit/s (high quality). Remind that PoliTO setup allows four quality levels, as explained in Section~\ref{sec:dataset_vc}. In only 10\% of the cases, streams reach 300~kbit/s or higher, meaning that ultra-high quality video and screen sharing are seldom used. In all cases, bandwidth is not a major problem for BBB, and students need no more than 0.5-1.0~Mbit/s to enjoy a lecture.

At last, we consider session duration in Figure~\ref{fig:virtual_classroom_duration}. Half of the streams last less than 30 minutes, but it is hard to link this short duration to abandonment by students, as the server could reset multimedia streams for other reasons (e.g., the lecturer temporarily mutes the microphone, or takes a break during a class). Interestingly, we still notice two bumps at 1.5 and 3 hours, which are the typical duration of PoliTO classes.

\textbf{Take away:}
\textit{Live classrooms put a large workload on teaching servers with significant peaks on scheduled lecture times. Yet, bandwidth is not a major problem. Students need less than 1~Mbit/s downstream bandwidth to fully enjoy live lectures. Not-live teaching media are heavily consumed also during evenings and weekends.}

\section{Network performance}
\label{sec:net_perf}

We now evaluate network performance metrics to understand whether people accessing teaching material experience problems to obtain the content. We first investigate the performance by breaking down data according to the students' Internet Service Providers (ISPs). Then, we break down figures according to the geographical region from where students connect. In both cases, we map their ISPs and regions using the MaxMind datasets.\footnote{\url{https://www.maxmind.com/}} Results in this section are computed by considering the downloads of on-demand video or teaching material with TCP connections. Both contents are provided through HTTP bulk transfers and thus constitute a valid download performance test. 
To reduce noise, we consider only download of objects of at least 10~MB. 

\subsection{Internet Service Providers breakdown}

\begin{figure}[t]
    \begin{center}
        \begin{subfigure}{0.45\textwidth}
            \includegraphics[width=\columnwidth]{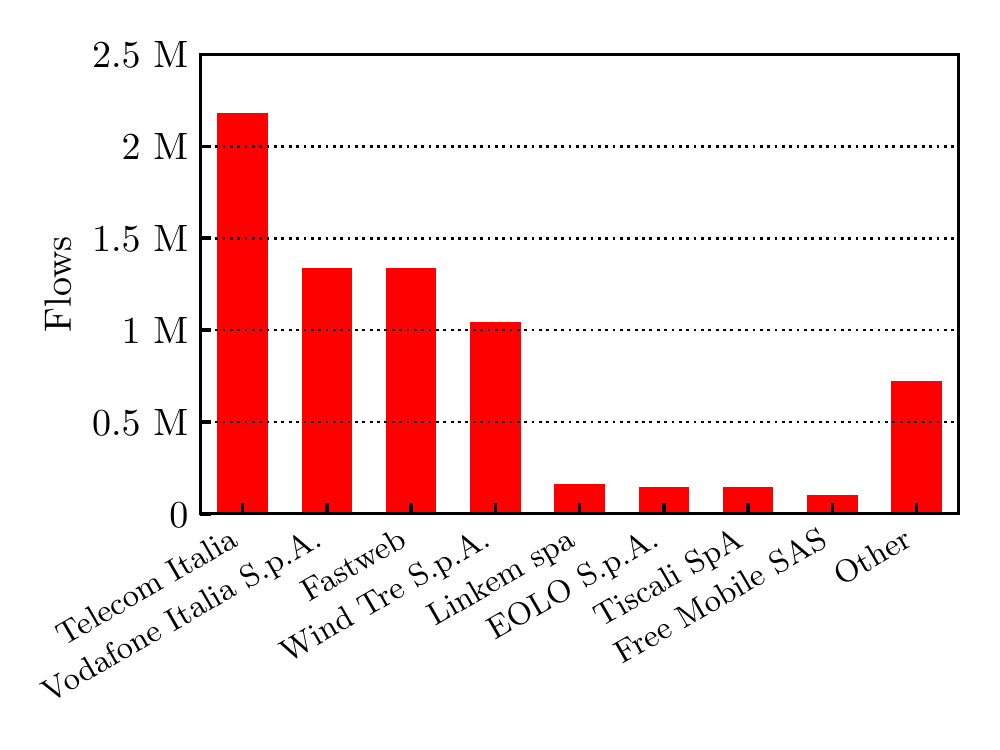}
            \caption{Number of download sessions}
            \label{fig:learning_ases_volume}
        \end{subfigure}
        \begin{subfigure}{0.45\textwidth}
            \includegraphics[width=\columnwidth]{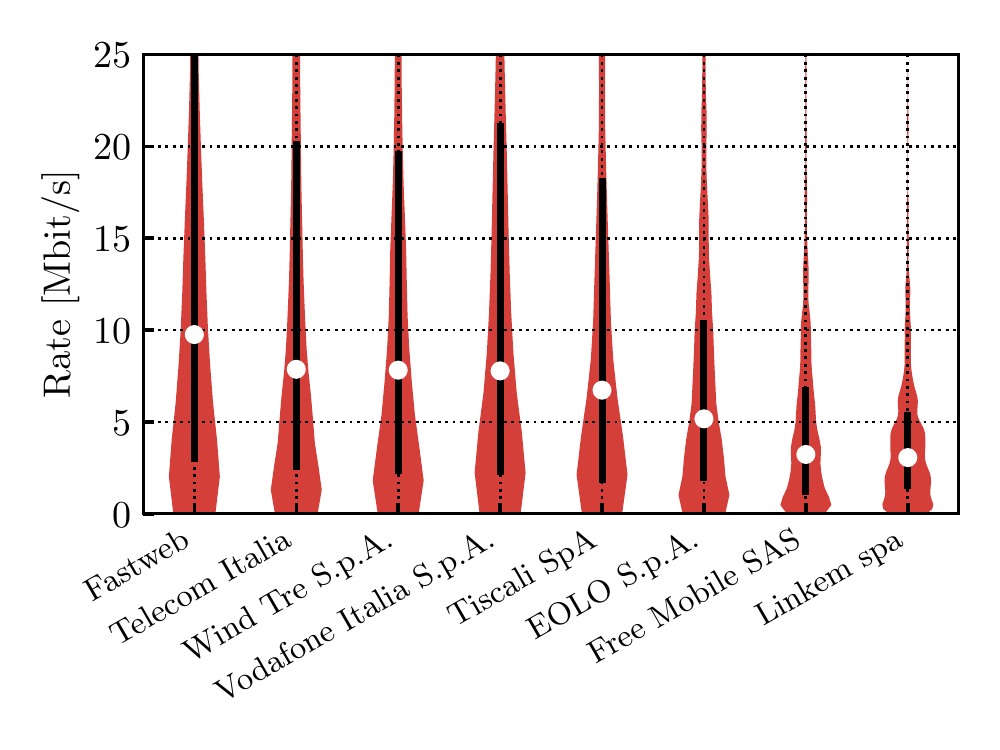}
            \caption{Download throughput distribution}
            \label{fig:learning_ases_throughput}
        \end{subfigure}
        \begin{subfigure}{0.45\textwidth}
            \includegraphics[width=\columnwidth]{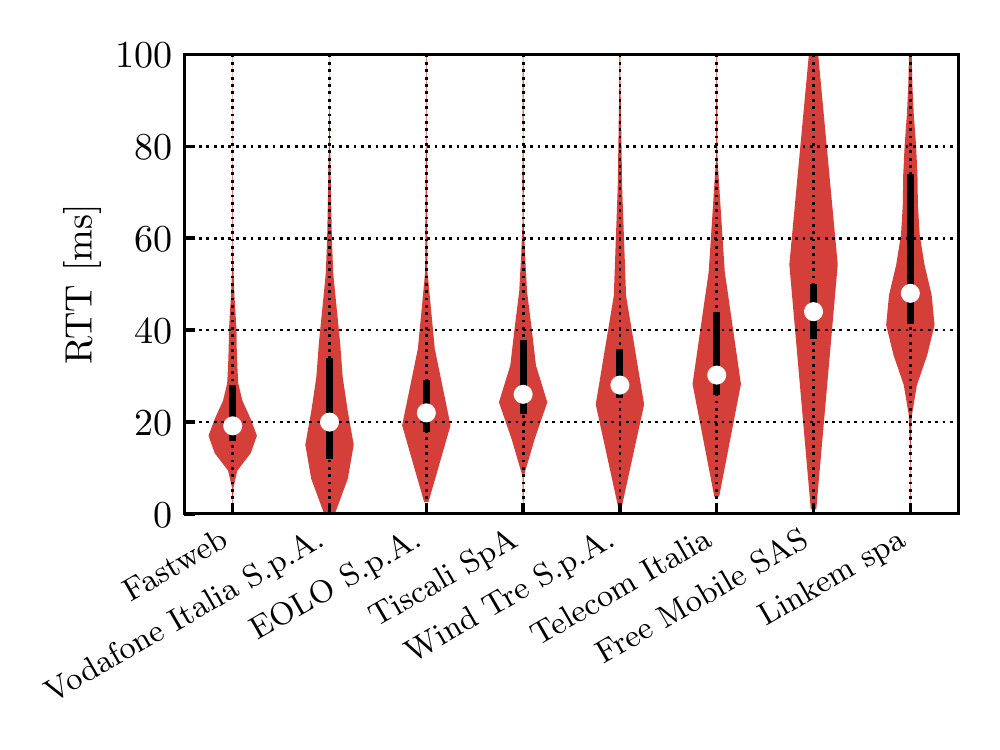}
            \caption{RTT distribution}
            \label{fig:learning_ases_rtt}
        \end{subfigure}
        \begin{subfigure}{0.4\textwidth}
            \includegraphics[width=\columnwidth]{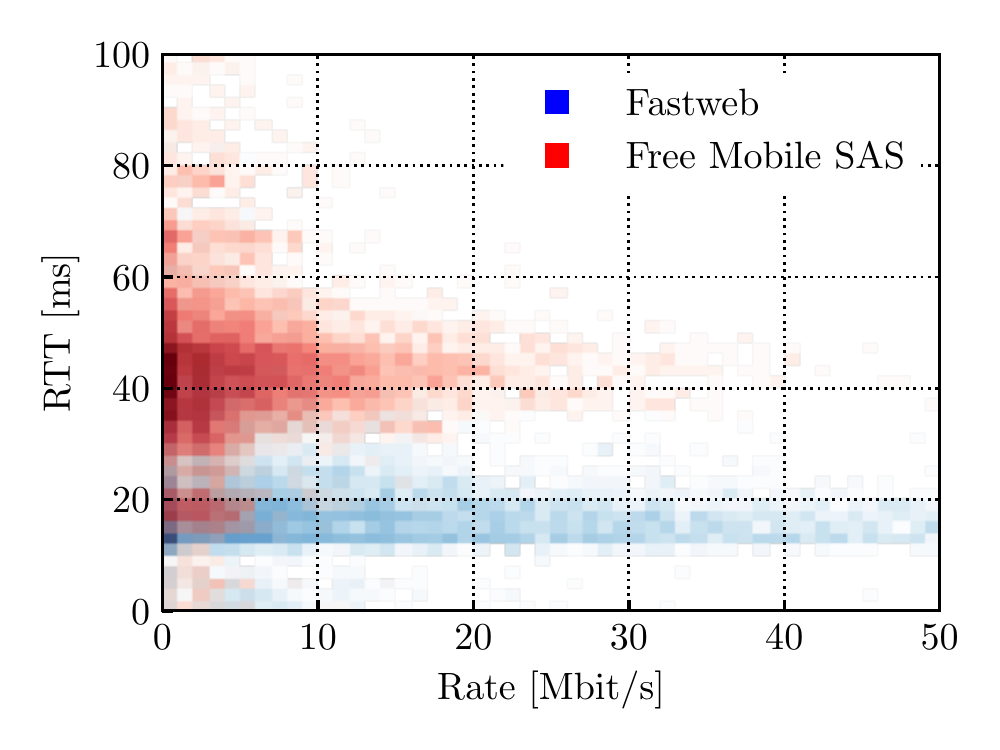}
            \caption{RTT vs throughput: Fastweb and Free}
            \label{fig:learning_ases_fw_free}
        \end{subfigure}
        \caption{Access to teaching servers per ISP. Only flows larger than 10~MB are considered.}
        \label{fig:learning_ases}
    \end{center}
\end{figure}

Figure~\ref{fig:learning_ases_volume} shows the distribution of the number of connections according to the ISPs hosting the client IP addresses. As expected, the largest 4 ISPs in Italy dominate the list, with TIM grabbing 31\% of the connections, followed by Vodafone, Fastweb and Wind Tre getting 20\%-15\% of the share.
Some ISPs rely on specific access technologies. For instance, EOLO, Linkem, and Free Mobile offer Internet over 3G/4G technologies only. Would students using them suffer eventual impairments?

For this, we check the distribution of download throughput for each ISP.
Violin plots in Figure~\ref{fig:learning_ases_throughput} depict results. Each violin plot represents the probability density function of the average per-flow throughput; White dots mark the median values of distributions. We sort providers by this median.
Most flows experience average throughput higher than 5~Mb/s. For example, the median on Fastweb customers (a provider offering mostly fiber access solution) is as high as 10~Mbit/s. Figures for 3G/4G-only providers are, instead, below 5~Mbit/s, thus pointing to possible client-side impairment (e.g., congestion).

Figure~\ref{fig:learning_ases_rtt} provides more insights. The violin plots depict the distributions of the minimum RTT which Tstat computes by measuring the time between a data segment and its corresponding acknowledgment. RTT is impacted (i) by the physical distance from clients to servers, (ii) by access technology delays and network congestion, and (iii) Internet routing.
Notice how the RTT distribution is very condensed for Fastweb (median at around 20~ms). On the other extreme, wireless-only providers have widespread distributions, with a median over 40~ms, and peaks above 100~ms (see Free Mobile as the clearest example). Figure~\ref{fig:learning_ases_fw_free} extends the analysis showing a per-flow comparison of RTT versus throughput for two providers. The low RTT for the (fiber-most) customers on Fastweb comes together with a large throughput. The more variable 3G/4G network on Free Mobile results in higher and more distributed RTT and lower throughput. In extreme cases, very large RTT values together with very low throughput are symptoms of possible network congestion. 

\textbf{Take away:}
\textit{Despite the variability of results, the very large majority of Italian ISPs can easily meet the roughly 1~Mbit/s required to enjoy live-streaming classes. Only a few customers, in particular those relying on 3G/4G access technologies, may suffer large delays and limited throughput with potential to disturb the user experience.}
 
\subsection{Geographical characteristics}

We repeat the performance analysis considering the geographical region from where students connect to the online teaching system. Students in some countries\footnote{\url{https://www.theguardian.com/commentisfree/2020/mar/23/us-students-are-being-asked-to-work-remotely-but-22-of-homes-dont-have-internet}} are reported to have problems following online lectures due to poor Internet connectivity. Whereas we cannot measure the number of students suffering from a total lack of connectivity, we can estimate whether PoliTO's students and faculty experience different performance when connecting from different regions of Italy.

\begin{figure}[t]
    \centering
    \includegraphics[width=0.35\textwidth]{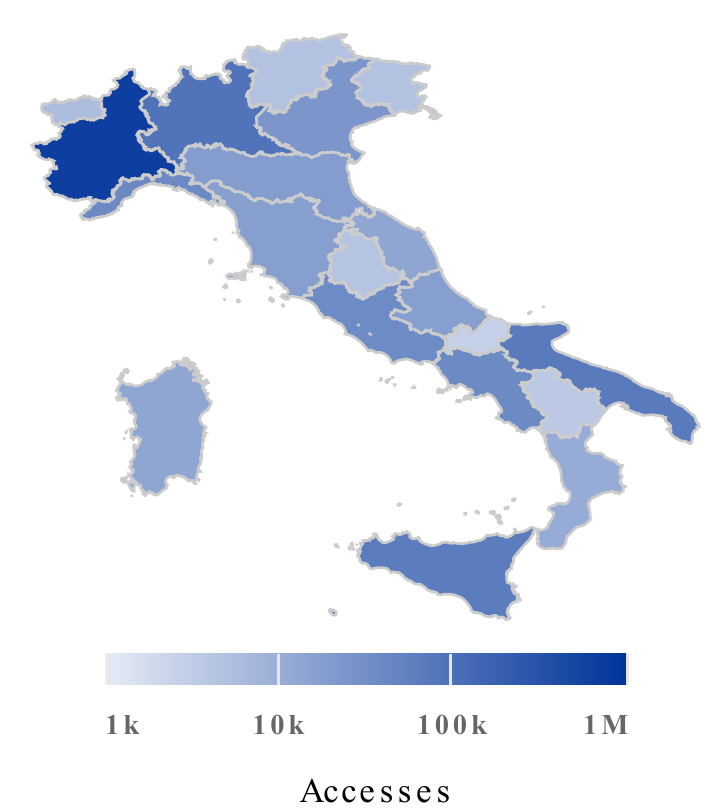}
    \caption{Access to teaching material per Italian region. Only flows larger than 10 MB are considered.}
    \label{fig:learning_regions_map}
\end{figure}

Figure~\ref{fig:learning_regions_map} visually reports the number of flows per Italian region. Again, we consider only connections downloading at least 10~MB. We can see that people have connected to the teaching system from all Italian regions. Naturally, a larger number of connections is seen for Piedmont, where PoliTO is located -- the region marked in dark-blue in northwestern Italy. Still, significant numbers of students are seen in other regions, allowing us to make fair comparisons.

\begin{figure}[!ht]
    \begin{center}
        \begin{subfigure}{0.7\textwidth}
            \includegraphics[width=\columnwidth]{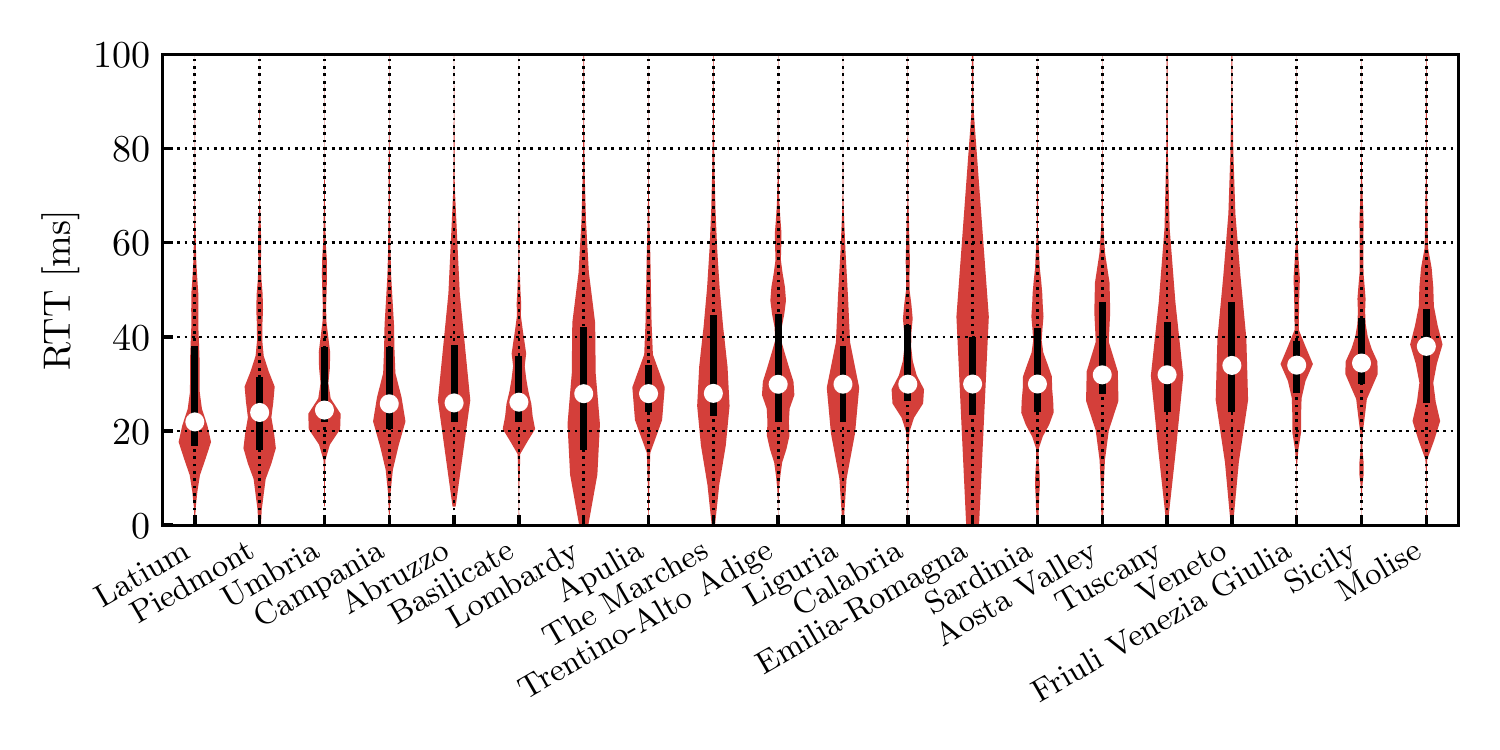}
            \caption{RTT distribution}
            \label{fig:learning_regions_rtt}
        \end{subfigure}\\
        \begin{subfigure}{0.7\textwidth}
            \includegraphics[width=\columnwidth]{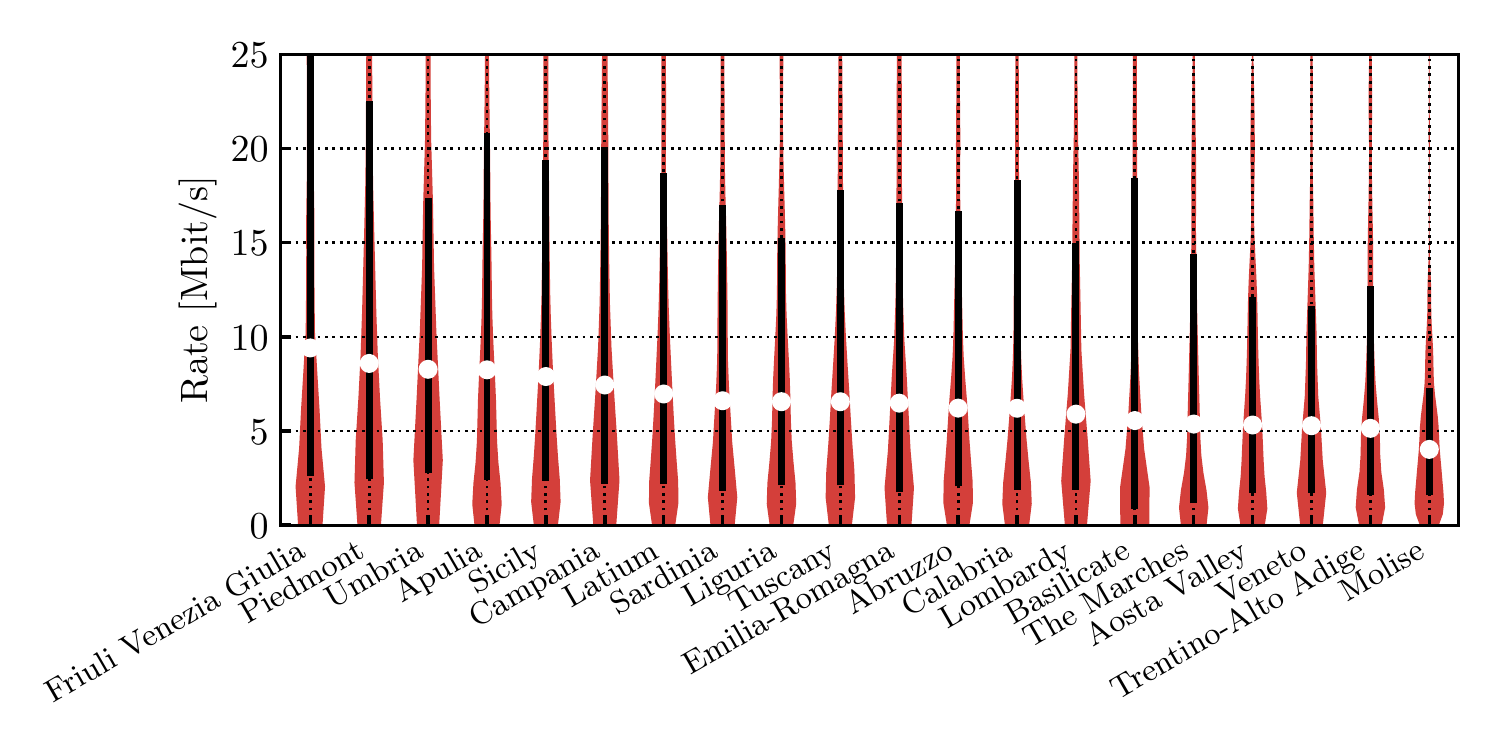}
            \caption{Download throughput distribution}
            \label{fig:learning_regions_rate}
        \end{subfigure}
        \caption{RTT and throughput distributions (flows larger than 10 MB) per Italian region.}
        \label{fig:learning_regions}        
    \end{center}
\end{figure}

Figure~\ref{fig:learning_regions} compares performance metrics for connections coming for the several Italian regions. We use violin plots once again, and sort regions in the $x-axes$ from the best to the worst median value for the metric in each plot.
Figure~\ref{fig:learning_regions_rtt} presents the distributions of RTT. Here physical distance is expected to play a key role. Yet, interesting patterns emerge. First, note that connections from Latium (the region where Rome is located) typically experience lower RTT than connections from Piedmont. This is an artifact due to Internet routing: Most Internet providers peer with GARR via Internet exchanges in Rome or Milan.
This aspect causes routing detours that artificially increase RTT for all regions, but Latium and in part Lombardy. Second, whereas southern regions are penalized by their distance to PoliTO, much noisier RTT distributions are seen for some northern regions. Observe, for example, that Veneto and Emilia-Romagna, two economically strong regions in northern Italian, have a large portion of connections with larger RTT than the median connection from Sicily (an island in southern Italy). This reflects again the different mix of access technologies and Internet routing of ISPs.

Figure~\ref{fig:learning_regions_rate} depicts distribution of the average connection throughput. We can see a large difference in throughput for the different regions. Connections coming from the best-performing regions have median average throughput twice as high as the regions with the lowest figures. Observe how some regions with large median RTT, such as Sicily and Friuli-Venezia-Giulia, still present good throughput figures. These results suggest that their large RTTs are mostly due to physical and routing distances rather than congestion. In some cases, large RTT variations, such as those observed for Veneto, are coupled with low throughput figures, which result in low Internet quality for a sizeable percentage of customers.

\textbf{Take away:}
\textit{While some people reported limited Internet performance due to traffic increase after COVID-19 lockdown~\cite{cloudflare,fastly}, our data show that overall performance is still good to access online teaching. Physical/routing distance has little impact in general, with Internet access technology that is still critical for flow performance.}

\section{Security events and unsolicited traffic}
\label{sec:unsolicited}

Finally, we check events visible from PoliTO's network security monitoring solutions. Our goal is to gauge evidence of possible changes in malicious network activities during the lockdown.

\begin{figure}[t]
    \begin{center}
        \includegraphics[width=0.7\columnwidth]{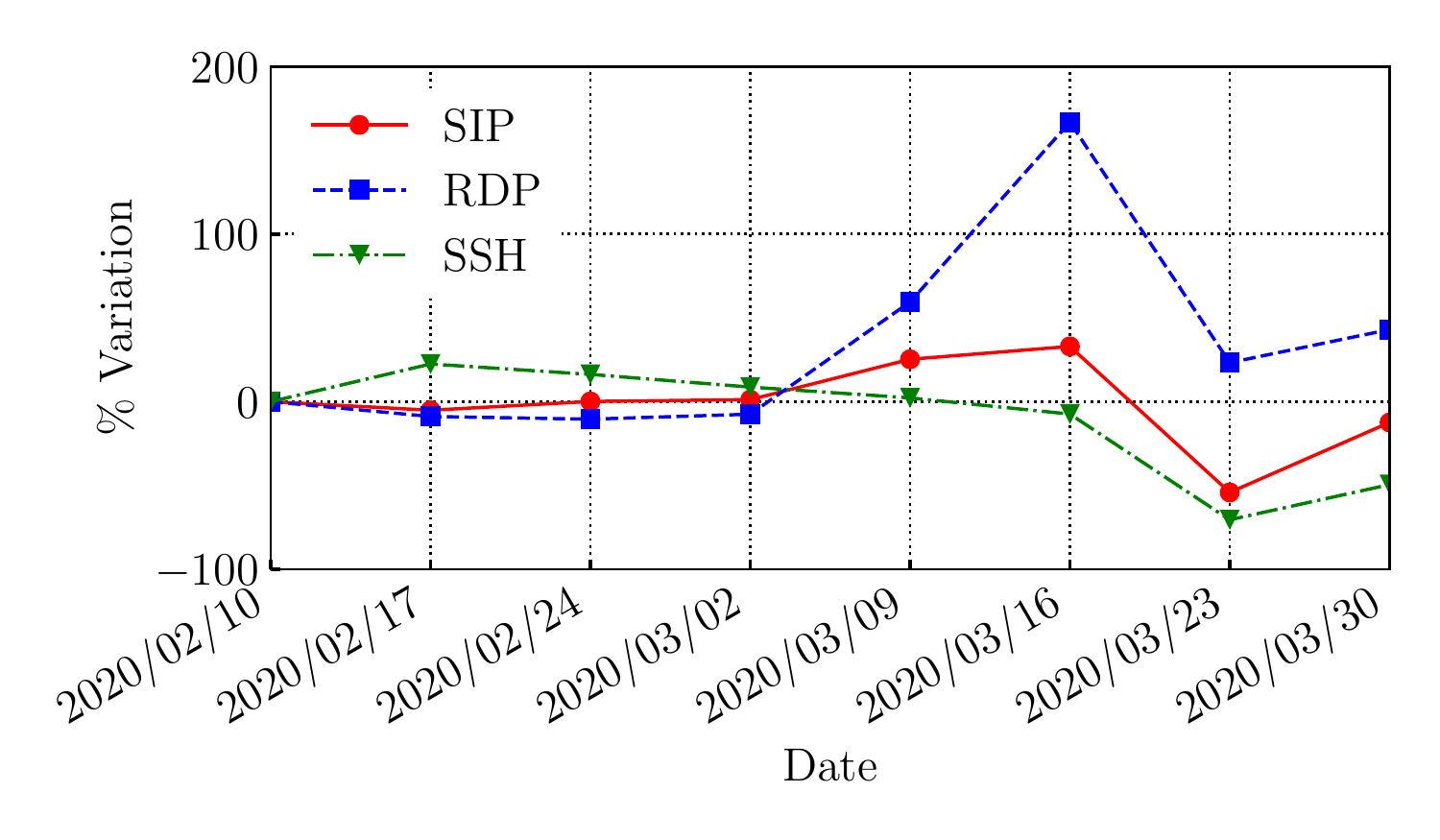}
        \caption{Variations in the number of events reported by the university firewall between February and March 2020.}
        \label{fig:IPS_1}
    \end{center}
\end{figure}

In Figure~\ref{fig:IPS_1} we report the variation on the number of events reported by the university firewall per week. The first week is taken as a reference. The three most common events are shown, namely scan/brute-force attacks against SIP, RDP and SSH. The numbers for SIP and SSH show little variation throughout the weeks. These small variations are in line with the normal operations of this firewall. For RDP, some significant changes appear in the week of March $16^{th}$. Indeed, the number of RDP events reported by the firewall has more than doubled in the first week of lockdown. Recall that many RDP servers were enabled during that week for easing smart working. Yet, this increase is a consequence of the deployment of the new systems and would happen even without the lockdown. As potential attackers have found new RDP servers online, they perform more scanning and brute-force activity against these nodes, too. The posterior decrease in RDP events is explained by changes in the firewall setup, performed to limit such scan activity. 

\begin{figure}[t]
    \begin{center}
        \includegraphics[width=0.7\columnwidth]{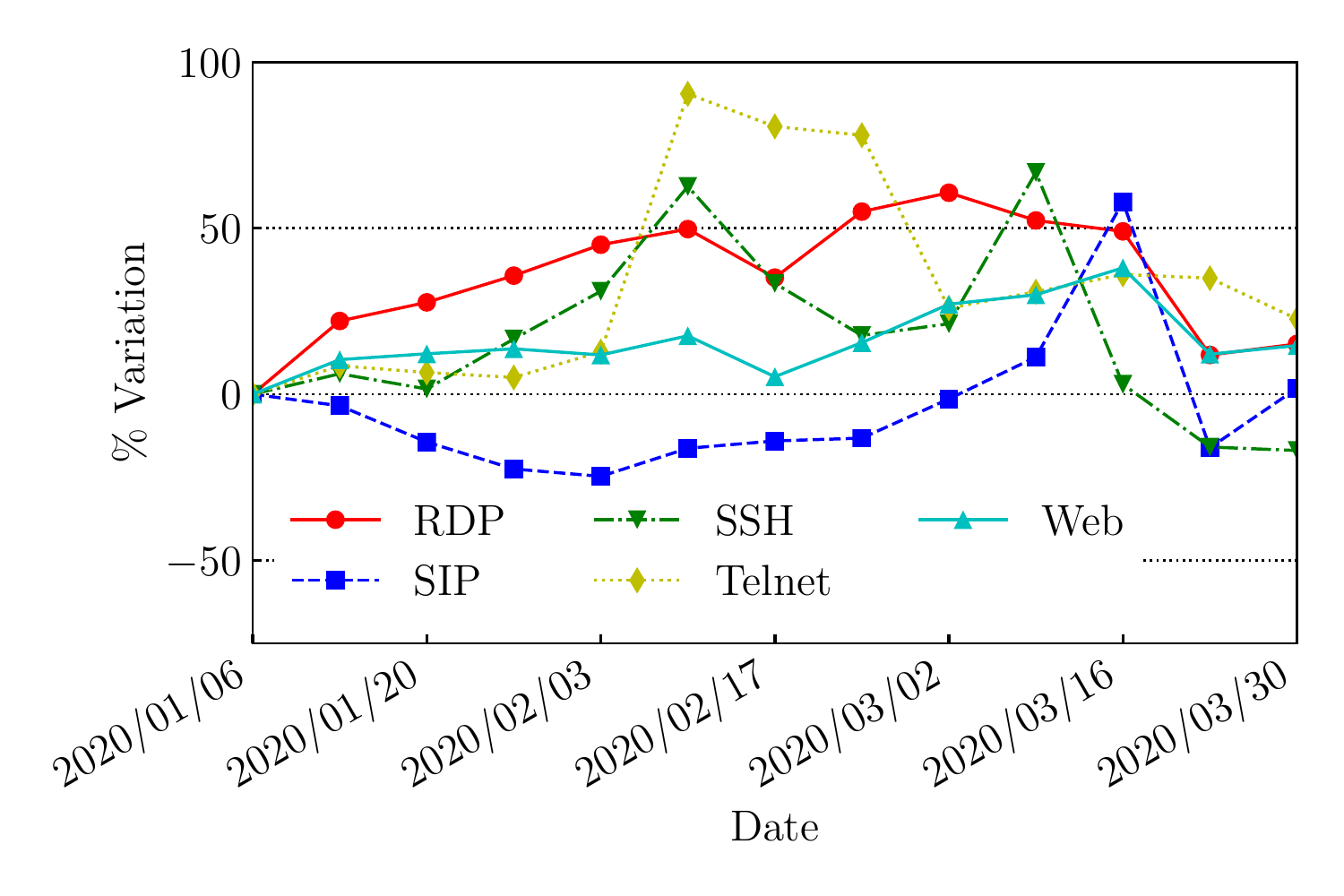}
        \caption{Variation in the darknet traffic between January and March 2020.}
        \label{fig:IPS_2}
    \end{center}
\end{figure}

A close inspection of traces of the darknet deployed at PoliTO leads to similar remarks. Figure~\ref{fig:IPS_2} shows the variations in the number of packets reaching the most contacted ports of the darknet. Traffic reaching the darknet is always noisy and great variations are normally seen when new large-scale Internet scans are performed~\cite{SoroDarknet2020}. We observe precisely this typical pattern during the weeks of lockdown as during any other period of time. We indeed observe episodes of sudden increases for some particular ports (e.g., port TCP/23 used by Telnet). However, there is no evident correlation among these variations and the lockdown itself.

At last, we check the load on email servers, but we omit figures for brevity. We observe a clear decreasing trend in SPAM emails. While the trend seems more prominent during March, the decrease has started months before. Again, there is no evidence that the reduction in SPAM would be a consequence of the lockdown.

\textbf{Take away:}
\textit{We observed little changes in unsolicited traffic, SPAM email and other security-related events. There are no clear signals that the most significant events are a consequence of the lockdown.}

\section{Related Work}
\label{sec:relatedwork}

The impact of the COVID-19 epidemic on the Internet ecosystem has been immediately measured by ISPs, content providers and specialized enterprises and published in the form of reports, news and blog posts.
SimilarWeb, a company that provides web analytics services for businesses, shows how the habits of people shifted during the outbreak~\cite{similarweb}, with some sectors like air travel, hotels and car rentals witnessing a large reduction of accesses while others (e-commerce, food delivery, and social networks) increased their popularity.
Microsoft reported a 755\% increase in usage of its cloud services~\cite{microsoft}, while YouTube and NetFlix reduced the streaming quality in Europe to prevent overload on the network~\cite{cnn}. CloudFlare and Fastly, two of the largest Content Delivery Networks worldwide, report a 20-40\% increase in daily traffic since the lockdown in Italy~\cite{cloudflare,fastly}, with somehow reduced performance.
Considering ISPs, the surge of network traffic has been testified with public news and blog articles. Vodafone states that fixed broadband usage has increased by more than 50\% in Italy and Spain~\cite{vodafone}, with a 100\% surge in upstream traffic and 44\% for downstream. Telefonica IP networks experienced a traffic increase of close to 40\% while mobile voice use increased by about 50\% and 25\% in the case of data~\cite{telefonica}. Also European Internet Exchange Points measured an increase of traffic in the order of 10-40\%~\cite{amsix,namex}.
Outside Europe, Comcast, one of the largest US operators, reports a 32\% increase in upstream traffic growth and an 18\% increase in downstream traffic growth~\cite{comcast}. In this work, we complement the above reports with additional figures showing the impact of the COVID-19 outbreak on the Campus networks and e-learning facilities.

Considering research papers, specific works targeting the effects of the COVID-19 pandemic have not yet been published. However, the role of the Internet has already been studied for the past catastrophic events.
Cho \emph{et~al.}~\cite{cho2011japan} study the impact of the 2011 earthquake in Japan on traffic and routing observed by a local ISP, while Liu \emph{et~al.}~\cite{cho2011japan} focus on inter-domain rerouting analyzing BGP data.
Zhuo \emph{et~al.}~\cite{zhuo2012egypt} underline the importance of the Internet during the 2011 Egyptian Revolt, while Groshek~\cite{groshek2012forecasting} finds statistical evidence that the Internet and mobile phones have helped to facilitate sociopolitical instability. Dainotti \emph{et~al.}~\cite{dainotti2011analysis} further address this aspect by analyzing the internet outages due to censorship actions that occurred during the Arab Spring revolts.
Heidemann \emph{et~al.}~\cite{heidemann2012preliminary} analyze network outages during the 2012 Sandy hurricane in the US. 
Sudden variations of Internet traffic have been observed also during massive software updates or following the born of new services (e.g., NetFlix) or protocols (e.g., Google QUIC)~\cite{finamore2011experiences,ruth2018first,trevisan2020five}.
Finally, some works propose general techniques to study or improve Internet traffic during disruptive events \cite{mackrell2013discovering, teixeira2006managing, liu2013characterizing}.
The events connected with the COVID-19 pandemic have a global scale and forced an unprecedented number of people to suddenly change their habits. As such, it is of great interest to study the impact on the Internet, and this paper provides an in-depth analysis of the Campus network traffic before and during the outbreak.

\section{Conclusions}
\label{sec:conclusions}

To mitigate the spread of the COVID-19 pandemic, the world issued severe restrictions like social distancing and lockdown measures. This forced people to change their habits and pushed them to online services for learning, smart working and leisure, generating an unprecedented load on the Internet. Since March 11$^{th}$ and still at the time of writing, Italy is facing a total lockdown, with 80-90\% of people forced to stay at home.

In this paper, we took the perspective of the Politecnico di Torino campus, analyzing the changes in traffic patterns due to the lockdown measures and the switch to online collaboration and e-learning solutions. We observe that incoming traffic drastically decreased, while outgoing traffic has more than doubled to support online learning. 

Remote working and online collaboration exploded as well, with hundreds of staff members using the Microsoft Teams collaboration platform, VPN and remote desktop services to keep working from home.
Since then PoliTO's in-house e-teaching infrastructure is serving more than $600$ daily classes to more than $16\,000$ students, generating peaks of 1.5~Gbit/s of traffic.
We also looked for eventual impairment suffered by students using different ISPs, or connecting from different regions. Results show that the campus and the Internet have proved robust to successfully cope with challenges and maintain the university operations.

\section*{Acknowledgements}
\noindent
The research leading to these results has been funded by the Smart-Data@PoliTO center for Big Data technologies.
We would like to thank PoliTO's IT Department ``Area IT'' and in particular Enrico Venuto and Giorgio Santiano for providing the data and precious support for the analyses.

\bibliographystyle{elsarticle-num-names}
\bibliography{covid}

\end{document}